\documentclass[12pt]{article}

\usepackage{amsmath,amsfonts,amssymb}

\usepackage{subfigure}

\usepackage{epsfig,graphicx,bm}

\usepackage{epstopdf}

\usepackage{lipsum}

\begin{document}
	
	\begin{titlepage}
		
		\begin{center}
			
			\vskip 0.4 cm

				{\Large \bf Note About Relational Mechanics of
				General Forms of Particle Actions}

			\vskip 1cm
			
			\vspace{1em}J. Kluso\v{n},		
			\footnote{Email addresses:	 klu@physics.muni.cz   (J.Kluso\v{n}), }\\
			
			\vspace{1em}
\textit{Department of Theoretical Physics and
				Astrophysics, Faculty of Science,\\
				Masaryk University, Kotl\'a\v{r}sk\'a 2, 611 37, Brno, Czech
				Republic}
			
			\vskip 0.8cm
			
		\end{center}
		
		\begin{abstract}
In this short note we show that any action for $N$ interacting
particles can be made invariant under gauged Galilean transformations. While resulting Lagrangian is generally very complicated its Hamiltonian has simple form  with first class constraints which are generators of corresponding gauge transformations included.

		\end{abstract}
	\end{titlepage}
	
	\bigskip
	
	\newpage

\def\bn{\mathbf{n}}
\newcommand{\bC}{\mathbf{C}}
\newcommand{\bD}{\mathbf{D}}
\def\hf{\hat{f}}
\def\tK{\tilde{K}}
\def\bp{\mathbf{p}}
\def\mC{\mathcal{C}}
\def\bJ{\mathbf{J}}
\def\bL{\mathbf{L}}
\def\bk{\mathbf{k}}
\def\tr{\mathrm{tr}\, }
\def\tmH{\tilde{\mH}}
\def\tY{\mathcal{Y}}
\def\nn{\nonumber \\}
\def\bI{\mathbf{I}}
\def\tmV{\tilde{\mV}}
\def\e{\mathrm{e}}
\def\bP{\mathbf{P}}
\def\bE{\mathbf{E}}
\def\balpha{\mathbf{\alpha}}
\def\bX{\mathbf{X}}
\def\bY{\mathbf{Y}}
\def\bR{\bar{R}}
\def\hN{\hat{N}}
\def\hK{\hat{K}}
\def\hnabla{\hat{\nabla}}
\def\hc{\hat{c}}
\def\mH{\mathcal{H}}
\def \Gi{\left(G^{-1}\right)}
\def\hZ{\hat{Z}}
\def\bz{\mathbf{z}}
\def\br{\mathbf{r}}
\def\mM{\mathcal{M}}
\def\vlambda{\vec{\lambda}}
\def\bK{\mathbf{K}}
\def\iD{\left(D^{-1}\right)}
\def\tmJ{\tilde{\mathcal{J}}}
\def\tr{\mathrm{Tr}}
\def\mJ{\mathcal{J}}
\def\mB{\mathcal{B}}
\def\partt{\partial_t}
\def\parts{\partial_\sigma}
\def\bG{\mathbf{G}}
\def\str{\mathrm{Str}}
\def\Pf{\mathrm{Pf}}
\def\bM{\mathbf{M}}
\def\tA{\tilde{A}}
\newcommand{\mW}{\mathcal{W}}
\def\bx{\mathbf{x}}
\def\by{\mathbf{y}}
\def \mD{\mathcal{D}}
\newcommand{\tZ}{\tilde{Z}}
\newcommand{\tW}{\tilde{W}}
\newcommand{\tmD}{\tilde{\mathcal{D}}}
\newcommand{\tN}{\tilde{N}}
\newcommand{\hC}{\hat{C}}
\newcommand{\hg}{g}
\newcommand{\hX}{\hat{X}}
\newcommand{\bQ}{\mathbf{Q}}
\newcommand{\hd}{\hat{d}}
\newcommand{\tX}{\tilde{X}}
\newcommand{\calg}{\mathcal{G}}
\newcommand{\calgi}{\left(\calg^{-1}\right)}
\newcommand{\hsigma}{\hat{\sigma}}
\newcommand{\hx}{\hat{x}}
\newcommand{\tchi}{\tilde{\chi}}
\newcommand{\mA}{\mathcal{A}}
\newcommand{\ha}{\hat{a}}
\newcommand{\tB}{\tilde{B}}
\newcommand{\hrho}{\hat{\rho}}
\newcommand{\hh}{\hat{h}}
\newcommand{\homega}{\hat{\omega}}
\newcommand{\mK}{\mathcal{K}}
\newcommand{\hmK}{\hat{\mK}}
\newcommand{\hA}{\hat{A}}
\newcommand{\mF}{\mathcal{F}}
\newcommand{\hmF}{\hat{\mF}}
\newcommand{\tk}{\tilde{k}}
\newcommand{\hQ}{\hat{Q}}
\newcommand{\mU}{\mathcal{U}}
\newcommand{\hPhi}{\hat{\Phi}}
\newcommand{\hPi}{\hat{\Pi}}
\newcommand{\hD}{\hat{D}}
\newcommand{\hb}{\hat{b}}
\def\I{\mathbf{I}}
\def\tW{\tilde{W}}
\newcommand{\tD}{\tilde{D}}
\newcommand{\mG}{\mathcal{G}}
\def\IT{\I_{\Phi,\Phi',T}}
\def \cit{\IT^{\dag}}
\newcommand{\hk}{\hat{k}}
\def \cdt{\overline{\tilde{D}T}}
\def \dt{\tilde{D}T}
\def\bra #1{\left<#1\right|}
\def\ket #1{\left|#1\right>}
\def\mV{\mathcal{V}}
\def\Xn #1{X^{(#1)}}
\newcommand{\Xni}[2] {X^{(#1)#2}}
\newcommand{\bAn}[1] {\mathbf{A}^{(#1)}}
\def \bAi{\left(\mathbf{A}^{-1}\right)}
\newcommand{\bAni}[1]
{\left(\mathbf{A}_{(#1)}^{-1}\right)}
\def \bA{\mathbf{A}}
\newcommand{\bT}{\mathbf{T}}
\def\bmR{\bar{\mR}}
\newcommand{\mL}{\mathcal{L}}
\newcommand{\mbQ}{\mathbf{Q}}
\def\mat{\tilde{\mathbf{a}}}
\def\mtF{\tilde{\mathcal{F}}}
\def \tZ{\tilde{Z}}
\def\mtC{\tilde{C}}
\def \tY{\tilde{Y}}
\def\pb #1{\left\{#1\right\}}
\newcommand{\E}[3]{E_{(#1)#2}^{ \quad #3}}
\newcommand{\p}[1]{p_{(#1)}}
\newcommand{\hEn}[3]{\hat{E}_{(#1)#2}^{ \quad #3}}
\def\mbPhi{\mathbf{\Phi}}
\def\tg{\tilde{g}}
\newcommand{\phys}{\mathrm{phys}}

\section{Introduction and Summary}
Relational mechanics  is closely 
related to Mach's idea that claims that dynamics of $N$ particles should be a theory 
of relations about these quantities without any reference to an external non-material entities.
Even if such a proposal is very intuitive it is not completely clear how  to make these ideas more concrete. One such a possibility
was proposal that the Lagrangian should be invariant under gauged Galilean group 
\cite{Barbour:1982gha} \footnote{See also \cite{Barbour:2010dp,Lynden-Bell:1995cmj}.}. Such Lagrangian can be found when  its measure (kinetic  energy term) is replaced  
with a measure that is defined in the space of orbits, where the orbits correspond to a set of configurations which are  equivalent under gauge transformations. Then it was shown that 
solutions of this gauge invariant dynamics correspond to the solutions of the original Lagrangian
with vanishing total momentum and angular momentum. An alternative proposal how to construct relational mechanics was presented in \cite{Ferraro:2014yza} where the starting point was  Lagrangian invariant under rigid Galilean transformation which is quadratic in velocities.  Then the variation of the kinetic
term under time dependent Galilean transformations leads to terms which are proportional to time derivative of parameters of these transformations. It was argued in 
\cite{Ferraro:2014yza} that when we add specific conterterms to the Lagrangian it is possible to find
Lagrangian which is invariant under  time dependent Galilean transformation
\cite{Ferraro:2014yza}.

Let us be now more specific.  Newton's mechanics is invariant under time independent translation, 
space translations and rotations of particle's positions $\bx_i$
\begin{eqnarray}\label{tr1}
&&	t'=t+\epsilon \ , \quad  \epsilon=\mathrm{const} \ , \nonumber \\
&&	\bx'_i=\bx_i+\vec{\xi} \ , \quad \xi=\mathrm{const} \ , \nonumber \\
&&	\bx'_i=\bA \bx_i \ , \nonumber \\
\end{eqnarray}
where $\bA$ is orthogonal matrix, $i=1,\dots,N$ where $i$ labels 
particles in ansamble. Further, Newton's laws are invariant under Galilean transformations
\begin{equation}\label{tr2}
\bx'_i=\bx_i+\mathbf{V} t \ , 
\end{equation}
which are special case of local time dependent translations when we identify $\xi=\mathbf{V}t$.
The transformations (\ref{tr1}) and (\ref{tr2}) represent Galilean group of Newtonian mechanics. 
In fact, there is a privileged set of inertial frames and clocks in Newtonian mechanics which are related each other through the transformation of the Galilean group. Then we leave an idea of these privileged frames when the Galilean group of transformation becomes gauge group with time dependent parameters.  In such a formulation of mechanics there are no privileged frames and clocks and 
it becomes purely relational. 


In this article we would like to generalize the procedure used in  \cite{Ferraro:2014yza} and in 
\cite{Glampedakis:2022fqu} to the case of collection of $N$ particles
where the kinetic term has square root structure and finally to the
case when the kinetic term has general form. This is very interesting
problem since for example an action for single D0-brane which is famous  Born-Infeld action \cite{Polchinski:1995mt,Dai:1989ua} has exactly this form. 
One can naively expect that since the action is non-linear in velocities it is impossible to perform gauging procedure as in 
\cite{Ferraro:2014yza}. However when we introduce  auxiliary modes  we can rewrite Lagrangian into the form which is quadratic
in velocities and then we can easily follow \cite{Ferraro:2014yza}. It is also clear that integrating out 
these auxiliary fields leads to complicated  form of the action. On the other hand when we proceed to the canonical formalism 
we find  Hamiltonian which is the same as in the case of the action invariant under rigid gauge transformation with 
the crucial difference which is the presence of six first class constraints 
\cite{Henneaux:1992ig}
that are generators of gauged Galilean transformations. Then we show that this result can be extended to the  
general form of kinetic terms. In other words, we show that any $N-$particle action with the potential 
term which depends on the relative distance of particles can be made invariant under gauged Galilean
transformation when the Hamiltonian has the same form as in the case of ungauged theory which is however supplemented by six first class constraints that are generators of gauged translations and rotations.

The structure of this paper is as follows. In the next section (\ref{second}) we find an action for $N$ particles
with kinetic terms that have square root structure 
that is invariant under gauged Galilean transformations. 
Then in section (\ref{third}) we determine its canonical form and identify constraints structure of the theory. Finally in section (\ref{fourth}) generalize this result to the case of arbitrary 
one particle actions.

\section{Action For $N$ Particles With Square Root Structure of Kinetic Term.}
\label{second}
In this section we consider  Lagrangian for $N$ particles where the kinetic term has square root structure. Explicitly, we have Lagrangian in the form  
\begin{equation}\label{Lagsquareroot}
	L=-\sum_i m_i\sqrt{1-\dot{\bx}_i^2}-V(\bx_{ij}) \ ,
\end{equation}
where $\Sigma_i\equiv \Sigma_{i=1}^N, \ m_i$ is a  mass of an individual particle with coordinates $\bx_i\equiv x^\alpha,\alpha=1,2,3$ and the potential depends on the relative distances of $i-$th and $j-$th particles 
\begin{equation}
	\bx_{ij}=\bx_i-\bx_j \ . 
\end{equation}
In order to see that it is possible to make this Lagrangian (\ref{Lagsquareroot}) 
invariant under gauged Galilean transformation we introduce
an auxiliary mode  $e_i$ and replace each kinetic term in    
(\ref{Lagsquareroot}) 
with following expression 
\begin{equation}\label{defaux}
	\frac{m_i}{2}\left(\frac{1}{e_i}(1-
	\dot{\bx}_i^2)+e_i\right)
\end{equation}
that is equivalent to the original one when we solve equation of motion for $e_i$. Explicitly, 
from (\ref{defaux}) we derive following equations of motion for $e_i$ 
\begin{equation}
	-\frac{1}{e_i^2}(1-\dot{\bx}_i^2)+1=0 
\end{equation}
that has solution 
\begin{equation}
	e_i^2=1-\dot{\bx}_i^2  \ . 
\end{equation}
Inserting this result into (\ref{defaux}) we find that it is equal to the kinetic
term given in (\ref{Lagsquareroot}).  Hence we start with following 
Lagrangian 
\begin{eqnarray}\label{LagNfun}
&&	L=-\frac{1}{2}\sum_i m_i(\frac{1}{e_i}(1-\dot{\bx}_i^2)+e_i)-V(\bx_{ij})=\nonumber \\
&&	=\frac{1}{2}\sum_i\frac{m_i}{e_i}\dot{\bx_i}^2-V(\bx_{ij})-
	\frac{1}{2}\sum_i m_i(\frac{1}{e_i}+e_i) 
\end{eqnarray}
that is now quadratic in velocities and hence we can expect that
it can be made invariant under time dependent Galilean transformations. To see this let us 
 start with  time dependent translation 
\begin{equation}\label{xtr}
	\bx_i'=\bx_i+\vec{\xi}(t) \ .
\end{equation}
Clearly potential term is invariant under this translation since
it depends on $\bx_{ij}$ only however kinetic term transforms as
\begin{equation}\label{deltaLkin}
	\delta L_{kin}=\vec{\xi}\cdot\sum_i \mu_i \dot{\bx}_i  , \quad \mu_i\equiv \frac{m_i}{e_i} \ .
\end{equation}
Now, as in \cite{Ferraro:2014yza} 
we  add following term to the action 
\begin{equation}\label{contraterm}
	-\frac{1}{2\mM}(\sum_j \mu_j \dot{\bx}_j)^2  \ , \quad \mM=\sum_j \mu_j \ . 
\end{equation}
Then the variation of the contraterm (\ref{contraterm}) under the transformation
(\ref{xtr}) is equal to 
\begin{eqnarray}
	\delta_{\xi}(	-\frac{1}{2\mM}(\sum_j \mu_j \dot{\bx}_j)^2)=
	-\frac{1}{\mM}\sum_j \dot{\vec{\xi}} \cdot \mu_j \sum_k\mu_k\dot{\bx}_k=
	-\dot{\vec{\xi}} \sum_i \mu_i \cdot\dot{\bx}_i 
\end{eqnarray}
and hence we see that this variation exactly cancels the variation of the original
Lagrangian (\ref{deltaLkin}). Finally note that the original kinetic
term together with contraterm (\ref{contraterm}) can be written 
in manifest relational form 
\begin{eqnarray}\label{kingauged}
&&L_{kingaug}=	\frac{1}{4\mM}\sum_{i,j} \mu_i\mu_j \dot{\bx}_{ij}^2
	=\frac{1}{2}\sum_{i=1}^N\mu_i\dot{\bx}_i\dot{\bx}_i
	-\frac{1}{2\mM}(\sum_{i=1}^N\mu_i\dot{\bx}_i)^2 \ . 
	\nonumber \\
\end{eqnarray}
As the next step we proceed to time dependent  infinitesimal rotation that in three dimensions
has the form 
\begin{equation}\label{rottime}
	\br'_i=\br_i+\vec{\alpha} \times \br_i \ , 
\end{equation}
where $\vec{\alpha}$  is vector directed along the axis of rotation. Since potential term is invariant under rigid 
rotations by definition it is also clear that it is invariant under (\ref{rottime}). On the other hand since now 
$\vec{\alpha}$ depends on time the kinetic 
 term (\ref{kingauged})  transforms
as
\begin{equation}\label{rotkin}
	\delta L_{kingaug}=\dot{\vec{\alpha}}\cdot \bJ \ , \quad 
	\bJ=\frac{1}{2\mM}\sum_{i,j}\mu_i \mu_j \bx_{ij}\times \dot{\bx}_{ij} \ . 
\end{equation}
Then in  order to compensate (\ref{rotkin}) we should add additional term into Lagrangian which is quadratic in $\bJ$
\begin{equation}\label{comrot}
	-\frac{1}{2}\bJ \cdot \bM \cdot\bJ \ , 
\end{equation}
where $M_{\alpha\beta}=M_{\beta\alpha}$ is symmetric tensor that is function of $\bx_{ij}$ only so 
that it transforms under time dependent rotation in the same way as in the case of constant rotations. Explicitly, under time dependent rotation  the term (\ref{comrot}) transforms as
\begin{eqnarray}
	\delta (	\frac{1}{2}\bJ \cdot \bM  \cdot\bJ )=
	\delta \bJ \cdot \bM \cdot \bJ+\frac{1}{2}\bJ\cdot \delta \bM \cdot\bJ \ ,  \nonumber \\	
\end{eqnarray}
where
\begin{eqnarray}\label{deltaJ}
&&	\delta J^\alpha=\epsilon^{\alpha\beta\gamma}\alpha^\beta J^\gamma+
	\dot{\alpha}^\beta I^{\beta\alpha} \ , \nonumber \\
&&	I^{\alpha\beta}=\frac{1}{2\mM}\sum_{i,j}\mu_i\mu_j (\bx_{ij}^2 \delta^{\alpha\beta}-
	\bx_{ij}^\alpha \bx_{ij}^\beta) \ .  \nonumber \\
\end{eqnarray}
Note that we can write the first equation in (\ref{deltaJ}) in 
vector notation as 
\begin{eqnarray}\label{deltabJ}
&&	\delta \bJ
	=\vec{\alpha} \times \bJ+\dot{\vec{\alpha}} \cdot \bI \ , \quad \vec{\alpha}\cdot \bI\equiv
\dot{\alpha}^\beta I^{\beta\alpha} \ . 	
	 \nonumber \\
\end{eqnarray}
Note that under  rotation
(\ref{rottime})
 the tensor $I^{\alpha\beta}$ transforms as
\begin{eqnarray}
	\delta I^{\alpha\beta}
	=\epsilon^{\alpha\omega \delta}\alpha^\delta I^{\delta\beta}+\epsilon^{\beta\omega\delta}\alpha^\omega  I^{\alpha\delta}
\end{eqnarray}
or with the help of vector notation
\begin{equation}\label{deltabI}
	\delta \bI=-\vec{\alpha}\times \bI-
	\bI\times \vec{\alpha} \ .
\end{equation}
We see that it is natural to choose $M_{\alpha\beta}$ as matrix inverse to 
$I^{\alpha\beta}$ so that 
\begin{eqnarray}\label{deltabjI}
	\delta (\frac{1}{2}\bJ \cdot\bI^{-1}\cdot\bJ)=
	\delta \bJ \cdot \bI^{-1}\cdot\bJ+\frac{1}{2}\bJ \cdot \delta \bI^{-1}\bJ=\vec{\alpha}\cdot \bJ
	\nonumber \\
\end{eqnarray}
using the fact that the variation (\ref{deltabI}) exactly cancels 
rigid transformation of $\bJ$ given in (\ref{deltabJ}).
Then it is clear that  (\ref{deltabjI}) exactly cancels variation of kinetic term (\ref{rotkin})
 so that the Lagrangian that is invariant under gauged translation and rotation has the form 
\begin{equation}\label{Lfinalrel}
	L=\frac{1}{4\mM}\sum_{i,j}\mu_i\mu_j\dot{\bx}_{ij}\dot{\bx}_{ij}-\frac{1}{2}\bJ\cdot \bI^{-1}
	\cdot \bJ -V(\bx)-\frac{1}{2}\sum_i \left(\mu_i+\frac{m_i^2}{\mu_i}\right) \ .
\end{equation}
Formally this Lagrangian has the same form as Lagrangian found in \cite{Ferraro:2014yza}
with one crucial difference which is the fact that $\mu_i$ are not constants but are auxiliary 
modes that should be integrated out
\footnote{It is natural to used $\mu_i$ instead of $e_i$ where both sets of variables are equivalent. 
	For example, let us check it in case when $N=1$ when the Lagrangian is equal to 
	\[
		L=\frac{1}{2}\mu\dot{\bx}^2-\frac{1}{2}(\mu+\frac{m^2}{\mu}) \]. Then the  equation of motion for $\mu$ gives
\[		\dot{\bx}^2-1+\frac{m^2}{\mu^2}=0 
		\Rightarrow 	\ , \quad \mu=\frac{m}{\sqrt{1-\dot{\bx}^2}} \]
	that inserting back to the Lagrangian above gives 
	$L=-m\sqrt{1-\dot{\bx}^2}$.}. Even if these equations of motion for auxiliary modes can be solved
at least in principle it is clear from the structure of the Lagrangian where both $\bJ$ and $\bI$ depend on $\mu_i$ that the resulting Lagrangian is very complicated. However we show in the next section that corresponding
Hamiltonian has relatively simple form.
\section{Hamiltonian Formalism}\label{third}
In this section we would like to find Hamiltonian corresponding 
to the Lagrangian (\ref{Lfinalrel}). First of all we determine conjugate momenta 
\begin{eqnarray}\label{defbp}
	\bp_k=\frac{\delta L}{\delta \dot{\bx}_k}
	=\frac{1}{\mM}\mu_k\sum_j\mu_j\dot{\bx}_{kj}
	-\frac{1}{\mM}\mu_k\sum_j\mu_j (\bI^{-1}\cdot \bJ)
	\times 
	\bx_{kj} \ ,
	\nonumber \\
\end{eqnarray}
 where we used
\begin{eqnarray}
	\frac{\delta J^\alpha}{\delta \dot{x}^\beta_k}=
	\epsilon^{\alpha\gamma\beta}\frac{1}{\mM}\sum_j x^\gamma_{kj} \ . 
	\nonumber \\
\end{eqnarray}
Further, from (\ref{defbp}) 
we immediately find primary constraint of the theory in the form 
\begin{equation}
	\bP\equiv \sum_k\bp_k=0 \ . 
\end{equation}
In order to find another constraint let us multiply $\bp_k$ with $\bx_k\times$ and define
$\vec{\mJ}$ as 
\begin{eqnarray}
\vec{\mJ}\equiv \sum_k \bx_k\times \bp_k \ . 
\end{eqnarray}
Now using
\begin{eqnarray}
	\sum_{k,j}\frac{1}{\mM}\mu_k\mu_j\bx_k \times \dot{\bx}_{kj}
=\bJ
	\nonumber \\
\end{eqnarray}	
and
\begin{eqnarray}
&&	\sum_k \frac{\mu_k\mu_j}{\mM}\bx_k\times (\bI^{-1}\cdot \bJ)\times \bx_{kj}=
	\nonumber \\
&&	=\frac{1}{2\mM}\sum_{k,j}\mu_k\mu_j ((\bI^{-1}\cdot \bJ)(
	\bx_{kj}\cdot \bx_{kj}-\bx_{kj} (\bx_{kj}\cdot \bI^{-1}\cdot \bJ)))=\bI \cdot \bI^{-1}\cdot \bJ=\bJ\nonumber \\
\end{eqnarray}
we immediately find that there are another primary constraints
$\vec{\mJ}\approx 0$. Then, with the help of canonical Poisson brackets
\begin{equation}
	\pb{x_i^\alpha,p_j^\beta}=\delta_{ij}\delta^\alpha_\beta
\end{equation}
we get 
\begin{equation}
	\pb{P^\alpha,P^\beta}=0 \ , \quad \pb{P^\alpha,\mJ^\beta}
=\epsilon^{\alpha\beta\delta}P^\delta \ , \quad 	
	\pb{J^\alpha,J^\beta}=\epsilon^{\alpha\beta\gamma}J^\gamma 
\end{equation}
that implies that $P^\alpha\approx 0, \mJ^\beta\approx 0$ are first class constraints. 

Now using (\ref{defbp}) we obtain that Hamiltonian is equal to
 \begin{eqnarray}\label{Hdef}
&&	H=\sum_k \dot{\bx}_k\cdot\bp_k-L=
\nonumber \\
&&=	\frac{1}{4\mM}\sum_{i,j}\mu_i\mu_j\dot{\bx}_{ij}\cdot\dot{\bx}_{ij}-\frac{1}{2}\bJ\cdot \bI^{-1}
	\cdot \bJ +V(\bx_{ij})+\frac{1}{2}\sum_i (\mu_i+\frac{m_i^2}{\mu_i}) \ . 
	\nonumber \\
\end{eqnarray}
As the next step we use (\ref{defbp}) to express $\dot{\bx}_i$ as 
\begin{eqnarray}\label{dotbx}
	\dot{\bx}_k=\frac{1}{\mu_k}\bp_k+\frac{1}{\mM}\sum_j\mu_j\dot{\bx}_j
	+(\bI^{-1}\cdot \bJ)\times \bx_k-(\bI^{-1}\cdot \bJ)\times \mathbf{R} \ , \quad \mathbf{R}=\frac{1}{\mM}\sum_j \mu_j\bx_j
	\nonumber \\
\end{eqnarray}
so that
\begin{eqnarray}
	\frac{1}{4\mM}\sum_{i,j}\dot{\bx}_{ij}\dot{\bx}_{ij}=
	\sum_k \frac{1}{2\mu_k}\bp_k^2+\frac{1}{2}(\bJ\cdot \bI^{-1})\cdot \bI \cdot (\bI^{-1}\cdot \bJ)
	\nonumber \\
\end{eqnarray}
using the fact that $\bP\approx 0$ and $\vec{\mJ}\approx 0$. 
Inserting this result into (\ref{Hdef}) we obtain final form of 
the Hamiltonian
\begin{equation}\label{Hfin}
	H=\sum_k\frac{1}{2\mu_k}\bp_k^2+V(\bx_{ij})+\frac{1}{2}\sum_k \left(\mu_k+\frac{m_k^2}{\mu_k}\right)+\vlambda_p \cdot \bP+
	\vlambda_J\cdot  \vec{\mJ}\ ,
\end{equation}
where $\vlambda_p,\vlambda_J$ are Lagrange multipliers corresponding
to the primary constraints $\bP\approx 0, \vec{\mJ}\approx 0$. We see that the 
Hamiltonian (\ref{Hdef})  has the same form as the Hamiltonian that is invariant
under rigid Galilean transformations with crucial difference
that it contains six primary constraints $\bP\approx 0 \ ,
\vec{\mJ}\approx 0$. 

It is instructive to perform reverse Legendre transformation and derive
Lagrangian from the Hamiltonian (\ref{Hfin}). The equation of motion for $x_k^\alpha$ is 
\begin{equation}\label{dotx}
	\dot{x}^\alpha_k=\pb{x^\alpha_k,H}=\frac{1}{\mu_k}p_k^\alpha+\lambda^\alpha_p +\epsilon^{\alpha\gamma\omega}\lambda_\gamma x_\omega 
\end{equation}
so that Lagrangian is equal to 
\begin{equation}\label{Lbp}
	L=\sum_k\dot{\bx}_k \bp_k-H=
	\frac{1}{2}\sum_k \frac{1}{\mu_k}\bp_k^2
	-V(\bx)-\frac{1}{2}\sum_k \left(\mu_k+\frac{m_k^2}{\mu_k}\right) \ . 
\end{equation}
As the final step we should express $\bp_k$ as function of velocities. When we multiply (\ref{dotx}) with $\mu_k$, perform summation over $k$ and use the constraint $\bP\approx 0$ w obtain
\begin{eqnarray}\label{lambdapn}
	\sum_k \mu_k\dot{x}_k^\alpha=\mM \lambda^\alpha_p+\mM 
	\epsilon^{\alpha\gamma \omega}\lambda^\gamma_J R^\omega \ . 
	\nonumber \\
\end{eqnarray}
In the same way we multiply (\ref{dotx}) with $\mu_k\bx_k\times $, perform summation over $k$ and use the constraint $\vec{\mJ}\approx 0$ to get
\begin{eqnarray}\label{lambdajp}
	\sum_k \mu_k\bx_k\times \dot{\bx}_k=\mM\mathbf{R}\times \vec{\lambda}_p
	+\sum_k \mu_k (\bx_k\cdot\bx_k)\vec{\lambda}_J-\bx_k (\bx_k\cdot \vec{\lambda}_J) \ .  \nonumber \\
\end{eqnarray}
Finally we express $\vec{\lambda}_p$ from (\ref{lambdapn}) as
\begin{equation}\label{lambdapv1}
	\frac{1}{\mM}\sum_k\mu_k\dot{\bx}_k-\vec{\lambda}_J\times \mathbf{R}
	=\vec{\lambda}_p \ . 
\end{equation}
Inserting this result into (\ref{lambdajp}) we
obtain
\begin{eqnarray}
	\sum_k\mu_k (\bx_k-\mathbf{R})\times \dot{\bx}_k=
	\sum_k \mu_k [(\bx_k\cdot\bx_k)\vec{\lambda}_J
	-\bx_k (\bx_k\cdot\vec{\lambda}_J)-\mM(\mathbf{R}\cdot
	\mathbf{R})\vec{\lambda}_J+\mathbf{R} (\mathbf{R}\cdot \vec{\lambda}_J) 
	\nonumber \\
\end{eqnarray}
that can be written as
\begin{equation}\label{x}
	\bJ=\bI \cdot \vec{\lambda}_J  \ ,  
\end{equation}
where we used the fact that $\bJ$ can be written in an equivalent form 
\begin{equation}
	\bJ=\frac{1}{2\mM}\sum_{i,j}\mu_i\mu_j\bx_{ij}\times \dot{\bx}_{ij}=
	\sum_i \mu_i\bx_i\times \dot{\bx_i}
	-\mathbf{R}\times \sum_k \mu_k\dot{\bx}_k \ . 
\end{equation}
Finally from (\ref{x}) we   express $\vec{\lambda}_J$ as
\begin{equation}\label{vlambdapf}
	\vlambda_J=\bI^{-1}\cdot \bJ \ .
\end{equation}
Inserting this result into (\ref{lambdapv1}) we obtain
\begin{equation}\label{vlambdajf}
	 \vlambda_p=
	\frac{1}{\mM}\sum_k\mu_k\dot{\bx}_k-(\bI^{-1}\cdot \bJ)\times \mathbf{R} \ . 
\end{equation}
Finally inserting (\ref{vlambdapf}),(\ref{vlambdajf}) into (\ref{dotx}) we obtain  that $\bp_k$ is equal to 
\begin{eqnarray}
	\bp_k=
	\mu_k (\dot{\bx}_k-\frac{1}{\mM}\sum_k \mu_k\dot{\bx}_k-
	(\bI^{-1}\cdot \bJ)\times (\bx_k-\mathbf{R}) )
	\nonumber \\
\end{eqnarray}
that agrees with the expression (\ref{dotbx}). Inserting this result
into (\ref{Lbp}) we clearly obtain original Lagrangian 
(\ref{Lfinalrel}) when we should again stress the crucial difference 
between Lagrangian and Hamiltonian description of gauge invariant theory. Lagrangian depends on $\mu_i$ in very complicated way while
in the Hamiltonian description the situation is much simple. Explicitly, from (\ref{Hfin}) we obtain following equation of motion for $\mu_i$
\begin{equation}
	-\frac{1}{2\mu_k^2}\bp_k^2+\frac{1}{2}-\frac{1}{2}\frac{m_k^2}{\mu_k^2}=0
\end{equation}
that can be solved for $\mu_k^2$ as 	
	\begin{equation}
	\mu_k^2=m_k^2+\bp_k^2 \ . 
\end{equation}
Then inserting this result into (\ref{Hfin}) we obtain 
final form of Hamiltonian 
\begin{equation}
	H=\sum_k \sqrt{m_k^2+\bp_k^2}+V(\bx_{ij})+\vlambda_p \cdot  \bP+\vlambda_J \cdot 
	\vec{\mJ} \ 
\end{equation}
that contains six constraints $\bP\approx 0 \ , \vec{\mJ}\approx 0$. In the next section 
we generalize this analysis to the case of arbitrary one particle kinetic terms. 

\section{General Case}\label{fourth}
We would like to show that we can make theory gauge invariant even if it is given as collection of different one particle kinetic terms. 
Let us consider Lagrangian in the form
\begin{equation}
	L=\sum_i f_i(\dot{\bx}_i^2)-V(\bx_{ij}) \ , 
\end{equation}
where $f_i$ are general functions. Introducing auxiliary modes
$a_i$ and $b_i$ we can rewrite the Lagrangian given  above into the form 
\begin{equation}
	L=\sum_i [f_i(a_i)+b_i(\dot{\bx}_i^2-a_i)]-V(\bx_{ij})
\end{equation}
or  equivalently
\begin{equation}
	L=\sum_i b_i\dot{\bx}_i^2+\sum_i[f_i(a_i)-b_i a_i]-V(\bx_{ij}) \ . 
\end{equation}
We see that this Lagrangian has the same form as Lagrangian given 
in (\ref{LagNfun})
when we replace $\mu_i$ with $b_i$. Then we can immediately write  its  gauge invariant extension 
\begin{equation}
	L=\frac{1}{4\mB}\sum_{i,j}b_ib_j \dot{\bx}_{ij}\dot{\bx}_{ij}
	-\frac{1}{2}\bJ\cdot \bI^{-1}\cdot \bJ-V(\bx_{ij})
	+\sum_i[f_i(a_i)-b_ia_i]
\end{equation}
where
\begin{equation}
	\bJ=\frac{1}{2\mB}\sum_{i,j} b_ib_j\bx_{ij}\times 
	\dot{\bx}_{ij} \ , \quad  I^{\alpha\beta}=\frac{1}{2\mB}\sum_{i,j}b_ib_j
	(\bx^2_{ij}\delta^{\alpha\beta}-\bx_{ij}^\alpha \bx_{ij}^\beta) \ , \quad \mB=\sum_j b_j \ . 
\end{equation}
Again we see that the Lagrangian is manifestly non-local on the other hand when we switch to the Hamiltonian formulation we again determine six first class constraints $\bP\approx 0 \ , 
\vec{\mJ}\approx 0$ so that the Hamiltonian is equal to 
\begin{equation}
	H=\sum_i\frac{1}{b_i}\bp_i^2+V(\bx_{ij})-\sum_i[f_i(a_i)-b_i a_i]+\vec{\lambda}_p\cdot 
	\bP+\vlambda_J\cdot \vec{\mJ}+\sum_i (v^i p_i^a+u^ip_i^b) \ , 
\end{equation}
where $p_i^a\approx 0, p_i^b\approx 0$ are conjugate momenta to $a_i$ and $b_i$ respectively.
Now the requirement of the preservation of the primary constraints 
$p_i^a\approx 0, p_i^b\approx 0$ leads to 
\begin{eqnarray}
&&\dot{p}_i^a=\pb{p_i^a,H}=-f'_i(a_i)+b_i\equiv \Sigma_i^a\approx 0 \ , 
\nonumber \\
&&\dot{p}_i^b=\pb{p_i^b,H}=\frac{1}{b_i^2}\bp_i^2-a_i\equiv \Sigma_i^b\approx 0 \ , 
\nonumber \\
\end{eqnarray}
where $\Sigma_i^a, \Sigma_i^b$ are secondary constraints. Since
\begin{eqnarray}
&&	\pb{p_i^a,\Sigma_i^a}=f''_i(a_i) \ , \quad 
	\pb{p_i^a,\Sigma_i^b}=1 \ , \nonumber \\
&&\pb{p_i^b,\Sigma_i^a}=1 \ , \quad \pb{p_i^b,\Sigma_i^b}=\frac{2}{b_i^3}\bp_i^2 \  \nonumber \\	
\end{eqnarray}
we see that they are second class constraints that can be explicitly solved at least in principle. Note that since $\pb{\bp_i,\Sigma_j^a}=\pb{\bp_i,\Sigma_j^b}=0$ we find that Dirac brackets 
between $\bp_i,\bx_j$ are the same as Poisson brackets. Now  from $\Sigma_i^a$ we find $b_i=f'_i(a_i)$ that inserting into $\Sigma_i^b=0$ we get
\begin{equation}
\frac{1}{f'^2_i(a_i)}\bp_i^2=a_i \Rightarrow \bp_i^2=f'^2_i(a_i)a_i 
\end{equation}
Now we can presume that this equation can be solved for $a_i=\psi_i(\bp_i^2)$. Then inserting this result into Hamiltonian we find its final form 
\begin{equation}
	H=\sum_i \frac{1}{f'_i(\psi_i(\bp_i^2))}\bp_i^2-
	\sum_i [f_i(\psi_i(\bp_i^2))-f'_i(\psi_i(\bp_i^2))
\psi_i(\bp_i^2)]+V(\bx_{ij})+\vlambda_p\cdot \bP+\vlambda_J\cdot \vec{\mathcal{J}} \ . 
\end{equation}
We see that the resulting Hamiltonian is sum of kinetic terms for individual particles
which is in sharp contrast with the Lagrangian formulation.

We should once again stress the main result of our analysis. We showed that arbitrary 
Lagrangian for $N$ interacting particles where they interaction depends on relative 
positions only can be made invariant under time dependent Galilean transformations.

 {\bf Acknowledgment:}

The work of J.K. was
supported by the Grant Agency of the Czech Republic under the grant
GA26-22343S.

	\end{document}